\documentclass[12pt,preprint]{emulateapj}
\usepackage{graphics}

\def\gtorder{\mathrel{\raise.3ex\hbox{$>$}\mkern-14mu
             \lower0.6ex\hbox{$\sim$}}}
\def\ltorder{\mathrel{\raise.3ex\hbox{$<$}\mkern-14mu
             \lower0.6ex\hbox{$\sim$}}}

\def\Msun{\>{\rm M_{\odot}}}

\shorttitle{White Dwarf Planet Detection}
\shortauthors{Debes et al.}

\begin{document}
\title{Cool Customers in the Stellar Graveyard II: Limits to Substellar Objects
around nearby DAZ White Dwarfs}
\author{John H. Debes\altaffilmark{1}, Steinn Sigurdsson\altaffilmark{1},
Bruce E. Woodgate\altaffilmark{2}}

\altaffiltext{1}{Department of Astronomy \& Astrophysics, Pennsylvania State
University, University Park, PA 16802}
\altaffiltext{2}{NASA Goddard Space Flight Center, Greenbelt, MD 27710}

\begin{abstract}
Results from a concerted Hubble Space Telescope (HST) survey of nearby white
dwarfs for substellar objects is presented.  A total of 7 DAZ 
white dwarfs with distances of $<$ 50 pc had high contrast and high spatial
 resolution NICMOS coronagraphic images taken
to search for candidate substellar objects at separations $\ltorder$10\arcsec\
away.  Limits to unresolved companions are derived through analysis of 2MASS
photometry of the white dwarfs compared to expected fluxes based on the 
WDs effective temperature, distance, and gravity.  Our HST survey of seven DAZ white dwarfs identified candidate companions for four of the white dwarfs.  For three of these four, HST and ground-based second epoch observations showed the candidates to be background stars.  The fourth white dwarf, which is close to the galactic plane, has seven candidate companions at distances of 2\arcsec\ to 4\arcsec, which remain to be followed up.  We find that for four of the white dwarfs
we are sensitive to planetary companions $\gtorder$10 M$_{Jup}$.  For all of the targets, we are sensitive
to companions $>$ 18 M$_{Jup}$.
The lack of 
significant near infrared excesses for our targets
limits any kind of unresolved companions
present
 to be substellar.  In light of these results we make several
comments on the possibility of determining the origin of metals in the
atmospheres of these white dwarfs.
\end{abstract}   


\section{Introduction}

The last ten years have shown a surge of new discoveries about objects of
substellar mass.  Radial velocity surveys of main sequence K-F stars have found
few brown dwarf companions at separations of $<$3 AU, but a profusion
of planetary mass companions \citep{marcy00}.
  Large all sky-surveys, such as 2MASS and 
SDSS have found large numbers of free floating brown dwarfs \citep{burgasser03,hawley02}.  Low mass
substellar objects down to planetary mass have been discovered in young 
clusters such as $\sigma$ Orionis \citep[][and references therein]{lada03}.
  At the same time, imaging
surveys of nearby main sequence stars have found several substellar companions
thanks to high contrast imaging \citep[e.g.][]{forveille04}.
  One population of stars which still has little data are
 intermediate mass stars with masses between 1.5-8 M$_{\odot}$. 

 The reason
for the dearth of information around intermediate mass stars
 is two-fold.  First, the majority of searches for
planetary systems focus on Solar System analogues. Secondly, there are 
 technical limitations to searching for planets and brown dwarfs around 
main sequence
F-B stars.  Radial velocity surveys rely on a large number of narrow
absorption lines
in the stellar spectrum to achieve high precision velocity measurements \citep{
delfosse98,griffin00}.
  As the effective temperature of a star increases,
 metal line strengths decrease and
there are fewer lines for measurement.  Stars with higher masses have a 
correspondingly smaller reflex motion due to a low mass companion.   
Radial velocity surveys of G giant stars can probe higher mass stars, but there
are only two planetary candidates currently reported \citep{sato03,setiawan05}.  More massive stars have higher luminosities
 as well, making high contrast imaging more limited 
in its effectiveness if one is looking for the thermal emission from a 
companion.  Reflected light from substellar companions is most useful
within a few AU of a star and is negligible at larger distances \citep{burrows04}.  In most cases, searches focus on detecting the thermal radiation from a
substellar companion.
 
\begin{deluxetable*}{llccccccc}
\tablecolumns{9}
\tablewidth{0pc}
\tablecaption{\label{tab:targs} Properties of the Target White Dwarfs}
\tablehead{
\colhead{WD} & \colhead{Name} & \colhead{M$_f$\tablenotemark{a}} &  \colhead{T$_{eff}$} & \colhead{t$_{cool}$} & \colhead{D} & \colhead{M$_i$} & \colhead{t$_{cool}$+t$_{MS}$} & \colhead{References} \\
& & \colhead{($\Msun$)} & \colhead{(K)} & \colhead{(Gyr)} & \colhead{(pc)} & \colhead{($\Msun$)} & \colhead{(Gyr)} & }
\startdata
0208+396 & G 74-7 & 0.60 & 7310 & 1.4 & 17 & 2.1 & 3.2  & 1  \\
0243-026 & G 75-39 & 0.70 & 6820 & 2.3 & 21 & 3.2 & 2.8 & 1 \\
0245+541 & G 174-14 & 0.76 & 5280 & 6.9 & 10 & 4.6 & 7.2 & 1 \\
1257+278 & G 149-28 & 0.58 & 8540 & 0.9 & 34 & 1.7 & 3.3 & 1 \\
1337+701 & EG 102 & 0.57 & 20435 & 0.1 & 25 & 1.6 & 3.3 & 2,3 \\
1620-391 & CD-38$^\circ$10980 & 0.66 & 24406 & 0.1 & 12 & 3.1 & 0.7 & 4 \\
2326+049 & G 29-38 & 0.70 & 11820 & 0.6 & 14 & 3.7 & 1 & 2,5 \\
\enddata
\tablenotetext{a}{Values for M$_{f}$, T$_{eff}$, and
t$_{cool}$ were determined from listed references.  Distances derived from parallax measurements
compiled in (1).  If not available, 
(3) and (5) were used.  See Section \ref{s4.2} for the calculation of 
M$_i$ and the WDs' total ages.}
\tablerefs{(1) \citet{bergeron01} (2) \citet{liebert04} (3) \citet{hip}
(4) \citet{bragaglia95} (5) \citet{vanaltena95}}
\end{deluxetable*}

The effects of higher intrinsic luminosity are illustrated by noting the
sensitivity to substellar companions of the NICMOS instrument on HST.  
 High contrast imaging can achieve $\Delta H
\sim$10 at 1\arcsec\ on the NICMOS coronagraph with PSF subtraction, allowing
45 M$_{Jup}$ mass companions to be detected around a 1 Gyr solar mass star.  For an
A star with a mass of 2 M$_\odot$ at 1 Gyr, a 90 M$_{Jup}$ companion can be detected.  Finally, more massive stars are rarer in local space, forcing 
observations of young star forming regions at larger distances.

Recent images of young HAe/Be stars with circumstellar disks such as HD 141569,
HR 4796A, and AB Aurigae
 motivate a search for planets 
around higher mass stars \citep{weinberger99,rayjay99,grady99}.  Sub-mm 
observations of warped and clumpy disks around stars
such as Vega and Formalhut show
that planet formation may be vigorous for higher mass central stars \citep{holland98}.
What is still unclear is how planet formation efficiency varies with
stellar mass and whether the brown dwarf desert is present over the same 
orbital separations for higher mass stars.

The 145 or so discoveries of planets by radial velocity surveys have told
us much about planet formation, but discoveries of planets in orbit around
post main sequence objects have the opportunity to challenge many accepted 
assumptions developed
on the basis of our current knowledge.  For example, the first terrestrial 
planets ever discovered were around a pulsar \citep{wolszczan92}.  The oldest Jovian planet
discovered in the M4 globular cluster in orbit around a white dwarf
 demonstrates that planet formation can occur in metal poor systems \citep{sigurdsson03}.
This discovery must be explained in the context of planet formation mechanisms
that favor stars with higher metallicity.

Detecting substellar companions in orbit around white dwarfs have several 
advantages compared to searching main sequence stars.  Given their intrinsic dimness, white dwarfs allow high 
contrast searches to probe interesting orbital separations \citep{burleigh02}.
  In 
addition, their higher effective temperature allows searches for unresolved
excesses at longer wavelengths \citep{ignace01}.  Given the range of
progenitor masses, white dwarfs probe a large
range of stellar mass.  Finally, they
complement radial velocity and transit searches that are biased towards close
companions.  High spatial resolution and high contrast imaging in the near
infrared with the NICMOS camera on HST allows the best chance for detecting
faint cool companions to nearby white dwarfs.  Planetary mass objects that are
less than 3~Gyr can be observed in the near-IR, specifically in the F110W ($\sim$J) and
F160W ($\sim$H) filters.  For example, a 3~Gyr old 10~M$_{Jup}$ planet can be observed
out to 20~pc with an HST observation of $\sim$1200s.

Searching a subset of white dwarfs that harbor markers for 
substellar objects can maximize the return of a survey.  Nearby hydrogen
white dwarfs with metal line absorption (DAZs) may fit this criterion.  Three
hypotheses have been put forth to explain the presence of DAZs: interstellar
matter (ISM) accretion \citep{dupuis92,dupuis93a,dupuis93b}, 
unseen companion wind accretion \citep{zuckerman03},
 and accretion of 
volatile poor planetesimals \citep{alcock86,debes02,jura03}.  

ISM accretion
has a wealth of problems in predicting many aspects of DAZs such as the large 
accretion rates required for some objects and the distribution of these objects
with respect to known clouds of dense material \citep{aannestad93,
zuckerman98,zuckerman03}.  The quick atmospheric 
settling times of hydrogen atmospheres imply that the white dwarfs are
 in close proximity with accretionary material.

Of the $\sim$34 DAZs known, seven of them have dM companions, supporting the argument that
 DAZs could have unseen companions that place material onto the 
WD surface through winds \citep{zuckerman03}.  In order to accrete enough material, 
 companions must be in extremely
close orbits, bringing into question why these objects have yet to be discovered
through transits, radial velocity surveys of compact objects, or
observable excesses in near-IR flux.  In most cases
the reflex motion from such objects would be easily detectable 
\citep{zuckerman92}.
  The idea of the presence of unseen companions 
also cannot explain objects like WD 2326+049 (G~29-38) which
 has an infrared 
excess due to a dust disk at roughly the tidal disruption radius \citep{graham90, patterson91}.

The invocation of cometary or asteroidal material as a method of polluting WD
atmospheres was developed to explain photospheric absorption lines due to metals in
the DAZ WD 0208+395 (G 74-7) \citep{alcock86}.  However, the rates
 predicted by these original studies could not
satisfactorily explain the highest accretion rates inferred for some objects
and could not easily reproduce the distribution of DAZs based on their 
effective temperatures \citep{zuckerman03}. 
However,
mixing length theory predicts a drop-off of observability for
accretion as a function of effective temperature which may swamp out 
the earlier prediction of \citet{alcock86} \citep{althaus98}.  
Also unclear is the 
effect non-axisymmetric mass
loss could have on the fraction of comet clouds lost by their hosts during
post main sequence evolution
\citep{parriott98}.  
By hypothesis, cometary clouds are the result of 
planet formation, so the long term evolution of planetary systems and their
 interaction with these comet clouds needs to be investigated 
\citep{tremaine92}.

The problems of the \citet{alcock86} model
 can be overcome by studying the stability of
planetary systems during evolution of the central 
star as it undergoes mass loss, leaving the main sequence and evolving
 into a white
dwarf.  Most planetary systems are stable on timescales 
comparable to their current age.
During adiabatic mass loss, companions expand their orbits
 in
a homologous way, increasing their orbital semi-major axes
 by a factor M$_i$/M$_f$ \citep{jeans24}.  This change in the central 
stellar mass affects the dynamics of the planetary system.

 The change in stellar mass specifically affects the stability planetary 
systems, typified by the
Hill stability criterion against close approaches for two comparable mass
 planets.  The stability criterion is
roughly described as $\Delta_c=(a_1-a_2)/a_1=3\mu^{1/3}$, where
$a$ is the semi-major axis, $\mu$ is the mass ratio of the planets to 
the host star, and $\Delta_c$ represents the critical separation at which
the two planets become unstable to close approaches \citep{hill86,
gladman95}.  The critical
separation grows as the relative separation of the two planets stays the
same, resulting in marginally stable systems being tipped over the edge of 
stability.  This instability can lead to orbital rearrangements,
the ejection of one planet, and collisions \citep{ford01}.
These three events dramatically change the dynamical state
of the planetary system, leading to a fraction of systems that perturb
the surviving comet cloud and sending a shower of comets into the inner system
where they tidally disrupt, cause dust disks, and slowly settle onto the
WD surface.  This modification of the comet impact model can explain the
accretion rates needed for the highest abundances of Ca observed and
 the presence of infrared excesses around WDs \citep{debes02}.

For two of the three above explanations, unseen planetary or substellar objects
lurk in the glare of nearby white dwarfs with metal lines in their atmospheres.
DAZs represent a promising population for a search for cool objects in orbit around WDs.  If such 
companions can be detected, this will open an exciting chapter in the study of
extra-solar planets by presenting several objects that can be directly detected
and characterized, constraining a host of theoretical issues, such as
extra-solar planetary atmospheres and the long term evolution of 
Jovian planets.  Such observations in the stellar graveyard can support future
missions dedicated to the detection and characterization of terrestrial planets.  White dwarfs represent an intermediate step between our current 
technology and what is needed for observations made with the James Webb Space
Telescope (JWST) and the Terrestrial Planet Finder (TPF).  Coupled with the
possible marker of metal absorption, a sample of nearby stars easier to study
than main sequence stars guaranteed to have some sort of planetary system
could enhance the efficiency of such long term searches and  may provide
extra clues to the nature of planet formation.

To that end, we were motivated to search the seven brightest and closest DAZ
 white dwarfs with the 
NIC-2 coronograph on the NICMOS instrument of the Hubble Space Telescope (HST).
This search was part of the the Cycle 12 program 9834, completed over the course of 2003 and 2004 with 14 orbits.  The first results from this survey
focused on WD 2326+049 (G 29-38), a DAZ with an infared excess \citep[][hereafter DSW05]{debes05a}.  We present the observations we made in 
Section \ref{s1} and detail our data analysis in Section \ref{s2}.
We  present
candidate planetary and brown dwarf companions in Section \ref{s3} as well
as place limits on the types of candidates we could have detected in
Section \ref{s4}.  Finally, we discuss the implications of our work and
 lay out future possibilities in Section \ref{s5}.
\begin{deluxetable*}{lcccc}
\tablecolumns{5}
\tablewidth{0pc}
\tablecaption{\label{tab:obs} Table of HST Observations}
\tablehead{
\colhead{WD} & \colhead{Observation Group} & \colhead{Date \& Time (UT)} & 
\colhead{Integration Time} & \colhead{Filter}}
\startdata
0208+396 & N8Q320010 & 2003-09-15 19:42:00 & 575.877 & F110W \\
 & N8Q322010 & 2003-09-15 20:10:00 & 575.877 & F110W \\
0243-026 & N8Q322010 & 2003-09-18 18:11:00 & 575.877 & F110W \\
 & N8Q323010 & 2003-09-18 18:39:00 & 575.877 & F110W \\
0245+541 & N8Q318010 & 2003-08-26 21:11:00 & 575.877 & F110W \\
 & N8Q319010 & 2003-08-26 21:39:00 & 575.877 & F110W \\
 & N8Q368010 & 2004-10-24 07:45:00 & 575.877 & F110W \\
 & N8Q369010 & 2004-10-24 09:11:00 & 575.877 & F110W \\
1257+278 & N8Q316010 & 2004-02-18 11:12:00 & 575.877 & F110W \\
 & N8Q318010 & 2004-02-18 11:41:00 & 575.877 & F110W \\
1337+701 & N8Q302010 & 2003-12-01 17:09:00 & 25.918 & F205W \\
 & N8Q302011 & 2003-12-01 17:11:00 & 25.918 & F205W \\
 & N8Q302020 & 2003-12-01 17:20:00 & 21.930 & F160W \\
 & N8Q302030 & 2003-12-01 17:24:00 & 19.936 & F110W \\
 & N8Q308010 & 2004-02-05 21:40:00 & 575.877 & F110W \\
 & N8Q309010 & 2004-02-05 22:45:00 & 575.877 & F110W \\
 & N8Q310010 & 2004-02-05 23:13:00 & 575.877 & F160W \\
 & N8Q311010 & 2004-02-06 00:27:00 & 575.877 & F160W \\
1620-391 & N8Q303010 & 2003-09-07 06:12:00 & 23.924 & F205W \\
 & N8Q303011 & 2003-09-07 06:13:00 & 23.924 & F205W \\
 & N8Q303020 & 2003-09-07 06:22:00 & 17.942 & F160W \\
 & N8Q303030 & 2003-09-07 06:25:00 & 15.948 & F110W \\
 & N8Q312010 & 2004-03-08 03:22:00 & 575.877 & F110W \\
 & N8Q313010 & 2004-03-08 03:52:00 & 575.877 & F110W \\
 & N8Q314010 & 2004-03-08 05:00:00 & 575.877 & F160W \\
 & N8Q315010 & 2004-03-08 05:27:00 & 575.877 & F160W \\
2326+049 & N8Q301010 & 2003-10-20 10:07:00 & 17.942 & F205W \\
 & N8Q301011 & 2003-10-20 10:08:00 & 17.942 & F205W \\
 & N8Q301020 & 2003-10-20 10:15:00 & 11.960 & F160W \\
 & N8Q301030 & 2003-10-20 10:20:00 & 11.960 & F110W \\
 & N8Q304010 & 2003-09-14 19:31:00 & 575.877 & F110W \\
 & N8Q305010 & 2003-09-14 19:59:00 & 575.877 & F110W \\
 & N8Q306010 & 2003-09-14 21:07:00 & 575.877 & F160W \\
 & N8Q307010 & 2003-09-14 21:35:00 & 575.877 & F160W \\
\enddata
\end{deluxetable*}

\section{Observations}
\label{s1}
Only $\sim$34 DAZs are currently known to exist, since the detection of their weak
metal lines are difficult without a high signal-to-noise, high resolution
spectrograph \citep{zuckerman03}.
Six of the most promising DAZ white dwarfs discovered or confirmed in the 
\citet{zuckerman03} survey
 were targeted for observation 
with NICMOS and are listed in Table \ref{tab:targs}.  Our seventh target,
WD~1620-391, was chosen for the presence of circumstellar gas absorption 
features as well as photospheric absorption due to Si and C \citep{holberg95}.
  We chose these targets
based on the fact that these were the brightest and closest DAZs known.  
  Each target was observed with
 the NIC-2 coronagraph in the F110W
filter.  The most promising targets, WD~2326+049, WD~1337+701, and WD~1620-391
 were
imaged in the F160W filter as well.  With the exception of the newly 
discovered DAZ GD~362, both WD 2326+049 and WD~1337+701 have
the highest [Ca/H] abundances measured \citep{gianninas04}.  WD~1620-391 was 
chosen for extra observations
due to the presence of its circumstellar material.  
These three targets were also observed
without the coronagraph for shorter exposures in the F110W, F160W, and F205W
filters in an attempt to resolve 
any smaller structure or companions
 at separations $<$ 0.8\arcsec. Acquisition images were used for the other
targets.  Following the 
prescription of \citet{fraquelli04}, two coronagraphic
exposures of $\sim$600 s were taken
at two different spacecraft roll angles.  Each exposure was separated by 
a differential roll angle of 10$^{\circ}$. The differential roll angle between images
 limits the angular separation at which one can
detect a point source, requiring at least a two pixel
separation between the centroids of the
positive and negative conjugates to avoid the self
subtraction of any point source companions.  This requirement is tempered by
the need to spend most of the HST orbit observing the target and not rolling
the spacecraft.  For our observations, we concentrated on integration time and
chose a roll angle of 10$^{\circ}$, leading to an inner radius limit to extreme
high contrast imaging with self subtraction of 0.86\arcsec.

Table \ref{tab:obs} shows a log for all of the observations taken along with
the total exposure times and the filter used.  Each F110W observation was
designed to be sensitive enough to detect an object with m$_{F110W}\sim$23 with
a S/N of 10,
which for a 1 Gyr substellar object at 10 pc would correspond to a 
$\sim$5~M$_{Jup}$ planet.  For our seven targets, which range in age from 1 Gyr
to 7 Gyr and 10 to 34 pc, we are sensitive to 7-18 M$_{Jup}$ objects.

\begin{figure}
\plottwo{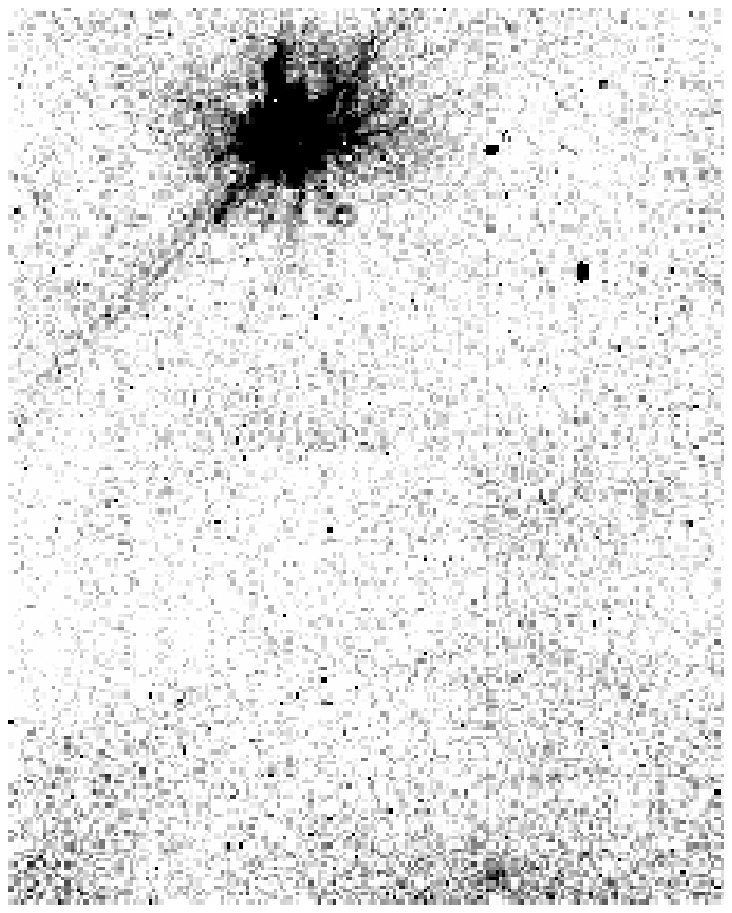}{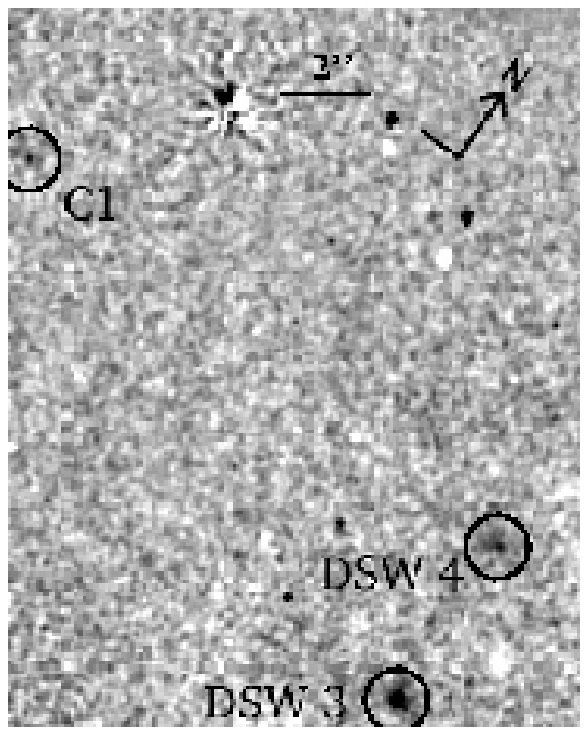}
\caption{\label{fig:g29fig} Image of WD 2326+049 in the F160W filter
before (left) and after (right) PSF subtraction.  The right panel has been
Gaussian smoothed to show a candidate and two extragalactic objects.  Other
dark features not marked are detector defects.}  
\end{figure}

In addition to the seven targets, three reference stars were imaged with
the three WDs observed without the coronagraph.  The goal was to use
these to subtract out the point spread function (PSF) that can obscure fainter 
objects or dust disks.  These targets were chosen to be close to the original 
target and have similar near-IR colors to aid in PSF subtraction.  No close
companions or structure were detected using the reference stars.

One group of observations taken of WD 0245+541 failed due to an incorrect 
calibration onboard the telescope, with the flight software (FSW). 
As a result, WD 0245+541 was not placed behind the coronagraphic hole.  The problem
was identified by the HST staff and further observations did not show the
same problem.  A repeat
observation was taken in October 2004, but the original failed
observations were also used for our data analysis.  

Due to the detection of a candiate planetary candidate around WD 2326+049 (G 29-38), second epoch observations were taken with the Gemini North Telescope 
using the Altair adaptive optics system in conjunction with the NIRI camera.
Altair can successfully guide on stars with
 R$\sim$13, such
as WD 2326+049.   By concentrating a diffraction limited
fraction of the total flux of a dim object, the background can be overcome for
extremely faint near infrared point sources.  In addition, under decent
observing conditions, the full width at half-maximum (FWHM) of
 the core on Altair images is $\sim$60-90 mas,
providing the possibility to resolve structures better than HST \citep{hutchings04}.

The Gemini observations were taken on August 5, 2004.
A total of 4 $\times$ 15s frames were co-added at 10 dither points to
subtract the background and to remove pixel to pixel defects, for an effective
integration on source of forty minutes.  Our total integration returned an
average FWHM of 75 mas, significantly smaller than the diffraction limit of
our F110W images with HST.

\section{Data Analysis}
\label{s2}
Data was reduced by the calibration pipeline provided for NICMOS.  In addition
to the pipeline, certain steps were taken in an effort to improve the quality
of the final images, roughly following the
procedure set out by \citet{fraquelli04}.
  Each 600 s exposure was broken up into two or
three exposures for ease in rejecting cosmic rays.  Each calibrated subexposure
had pedestal subtraction by the PEDSUB routine in IRAF through the STSDAS
package.  Each subexposure was registered and median combined with sigma 
clipping to create a final exposure at a particular roll angle.  The two
images at different roll angles were subtracted one from the other and vice
versa to create two difference images: a ROLL1-ROLL2 image and a  
ROLL2-ROLL1 image.  One difference image was rotationally registered and median
combined to produce the final total image.  Figure \ref{fig:g29fig} shows
the before and after pictures of a subtraction shown at the same image stretch.
The residual light due to the coronagraphic PSF is dominated by systematic 
errors, but in general is a factor of 20-50 times dimmer after subtraction.

In the case of WD 0245+541, several other steps had to be taken 
for the failed observation since at each roll
angle the star was at a different position and not behind the coronagraphic
hole.  To combat the poor positions, the two images were registered and 
difference images
were produced.  The final result was of sufficient quality to determine
the presence of several candidate objects in the field.

\section{Candidate Companions and Extragalactic Objects}
Of the seven targets, only four showed candidate companions in their fields.
The rest did not show anything with the exception of WD 1257+278, which had 
a resolved galaxy in the background.  Any extended objects were interpreted to
be background objects and all point sources were flagged as potential companions.  Where second epoch images were available with 2MASS or the POSS survey,
they were used or second observations were taken.  Each candidate with second
epoch images was checked for common proper motion with the target WD by measuring the relative radius and pointing angle in degrees East of North of the companion.  Extragalactic objects could potentially be of interest due to their
proximity to a bright object that could be used for guiding in a laser AO system
or multi-conjugate AO system.

The second epoch Gemini data was processed using several IRAF tasks designed by
the Gemini Observatory and based upon the sample scripts
 given to observers.  Each
frame was flatfielded and sky subtracted.  In addition, due to the on-sky rotation from the Cassegrain
 Rotator being fixed, each frame was rotationally registered
and combined.

To determine if an object had common proper motion with a target WD,
we calculated the predicted motion of the WD on the sky based on its proper
motion.  When comparing possible companions with 2MASS or POSS data, proper
motion alone was sufficient to determine objects that were in the background.
For WD~2326+049, WD~1620-391, and WD~0245+541, the annual parallactic
motion of the star was also calculated for
 an added means of determining background point
sources.  Any object in orbit around a WD would also have to share both 
proper motion and annual parallactic motion. 

 It is also important to 
adequately understand the errors in order to detect any possible proper 
motion of the background object or to determine how significant a measure of 
common proper motion is.  The greatest sources of error are due to 
uncertainties in the parallax of the WD, proper motion, and centroiding
errors in the PSF of the candidate.  Centroiding errors for faint sources
can be determined by looking at images in two filters for one of our fields
that has a lot of background sources.  The field of WD 1620-391 has several 
background point sources that can be compared between filters and two
epochs.  Comparing the difference of $\sim$30 sources between the F110W and 
F160W filters of the observation sets of N8Q312010 and N8Q314010 
yields a standard deviation between sources of $\sim$10 mas,
 which we will adopt as our general centroiding error. 

\label{s3}
\subsection{WD 2326+049}

Figure \ref{fig:g29fig} shows the NICMOS field of view around this WD, which included a 
candidate planetary companion, that we designated C1.  In addition,
there were two faint extended galaxies in the field. C1 
is discussed in more detail in DSW05 and
has been confirmed to be a background object with a second epoch observation
with the Gemini North Altair+NIRI instrument.  If
this object had been associated, its F110W and F160W magnitudes
 were consistent with
a 7 M$_{Jup}$ object.
  Table \ref{tab:gals}
 presents all the extragalactic objects discovered in this survey along
with their positions and apparent 
Vega magnitudes in the F110W and F160W filters
\begin{figure}
\plotone{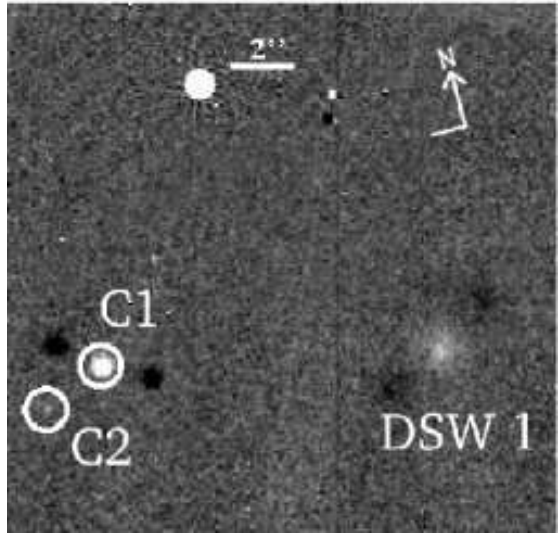}
\caption{\label{fig:g747} Field of WD 0208+395 with its candidates.  Candidates
are circled and the WD is masked to hide the systematic subtraction errors.
A galaxy is detected in the lower right of the image.}
\end{figure}

\subsection{WD 0208+395}

Figure \ref{fig:g747} shows two candidate objects, C1 and C2
 and a galaxy in the  field of WD~208+395.
Since the separation between these objects and WD~208+395 were greater than a few arcseconds, we pursued a 
second observation with the Canada France Hawaii Telescope with the PUEO+KIR
instruments.
A second epoch image shows that both C1 and C2
are in the background.  This result is discussed in detail in \citet{debes05c}.
  If they had been associated, C1 was consistent with
a 3 Gyr old 15 M$_{Jup}$ brown dwarf and C2 consistent with 
a 10 M$_{Jup}$ planet.

\begin{figure}
\plotone{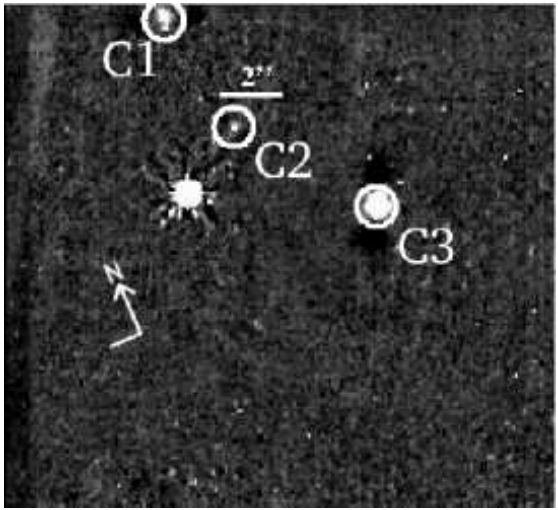}
\caption{\label{fig:g174} Field of WD 0245+541 with its candidates circled.}
\end{figure}

\subsection{WD 0245+541}
This object, due to its failed first observation, was re-imaged $\sim$ one year
later (see Table \ref{tab:targs}) which provided an ample baseline to test candidates for common proper
motion.
Figure \ref{fig:g174} shows the surrounding area of WD 0245+541, along with three
candidates in the field.  C1 appears to be a binary object at a distance
of $\sim$3\arcsec\, which in the second epoch image is clearly not co-moving.
  C2 is at a separation of $\sim$6\arcsec\
and 270$^\circ$ PA.
  Inspection of the POSS2 red image of this field clearly shows a point source
at a separation consistent with this object being a background source.
   Finally, C3 has a separation of 2.5$\pm0.02$\arcsec\ and 348$\pm$1$^\circ$ PA.  WD 0245+541 has a predicted motion between the two epochs of -667 mas
and -475 mas, leading to a predicted $\Delta\alpha$=0.17$\pm$0.12\arcsec\ and 
$\Delta\delta$=2.94$\pm$0.12\arcsec\ if C3 is non co-moving, compared to the
 observed $\Delta\alpha$=0.07 and
$\Delta\delta$=2.86.  The candidate does not have common proper motion, and is
therefore a background object.
 The main source of error was in the reported proper motion,
which had quoted errors of 0.1\arcsec yr$^{-1}$ \citep{bakos02}. 
If C3 had been associated, its F110W magnitude would have been
 consistent with an 18 M$_{Jup}$ brown dwarf
companion.  

\begin{deluxetable*}{cccccc}
\tablecolumns{6}
\tablewidth{0pc}
\tablecaption{\label{tab:gals} Table of Extragalactic Objects}
\tablehead{
\colhead{DSW \#} & \colhead{RA} & \colhead{Dec} & \colhead{F110W} & \colhead{F160W} & \colhead{Notes}
}
\startdata
 1 & 02 11 20.51 & +39 55 14 & 21.36$\pm$0.04 & &  \\
 2 & 12 59 45.63 & +27 34 01 & 22.8$\pm$0.1 & & $\sim$1.4\arcsec\ extent \\
 3 &23 28 47.96 & +05 14 38 & 23.7$\pm$0.2 & 22.1$\pm$0.1 & 0.23\arcsec\ aperture
 \\
4 & 23 28 47.67 & +05 14 40 & 24.0$\pm$0.2 & 22.8$\pm$0.2 & 0.23\arcsec\ aperture \\ 
\enddata
\end{deluxetable*}

\subsection{WD 1620-391}

\begin{figure}
\plotone{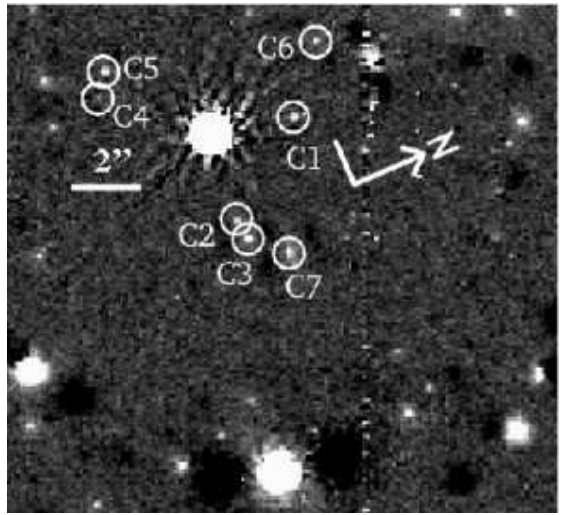}
\caption{\label{fig:wd1620} Field of WD 1620-391 with its candidates.  Each
candidate that is circled is within 4\arcsec\ and has colors consistent within
 the photometric errors to a candidate planetary object.}
\end{figure}

\begin{figure}
\plotone{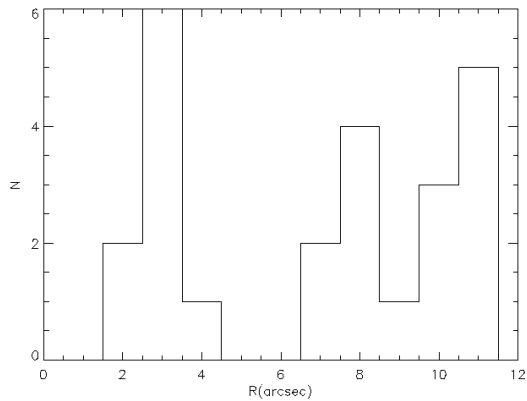}
\caption{\label{fig:dist} Distribution of point sources as a function of 
distance around the white dwarf WD 1620-391. }
\end{figure}

WD 1620-391 resides quite near the galactic plane and as such
has an extremely crowded field with $\sim$36 sources of varying brightnesses, 
which can be seen in Figure \ref{fig:wd1620}.
Any possible companion must
be separated from background objects.  A viable candidate
in this field would have to be selected by an F110W-F160W color
being consistent with a substellar object.  Since most of these objects are 
background objects we must first see if there is any evidence to suspect that
there would be a candidate in this field rather than assuming that all
sources were background objects.  The number of objects as a function of 
distance should be $\propto\ r^2$ if the background distribution is truly 
random.  A different distribution would be caused either by the presence of
objects physically associated to the central white dwarf or due to physical
associations among background stars, such as binaries or clustering.  To 
look for a departure from the expected distribution,
 we plotted the number of sources in the WD1620-391 field as a function of 
radial distance from the WD as shown in Figure \ref{fig:dist}. 
 We compared this distribution to a pure $r^2$ 
distribution through means of a K-S test.  We find that there is a 
97\% probability that the distribution is not based on the $r^2$ distribution
mainly due to the hump of sources present close to the WD.  We believe that 
those objects are viable candidates and that in a statistically 
significant way the distribution of sources $<$ 4\arcsec\ is fundamentally
different than what would be expected.  A caveat, however, is that since
the WD is at a low galactic latitude the statistical test may merely be 
detecting some fundamental structure in the background sources rather than
the presence of a candidate.  Additionally, the scenario of \citet{debes02} 
would predict more than one planet in the system to efficiently slingshot
comets or asteroids to the surface of the white dwarf, so the potential exists
that two planetary candidates could be present in this ``hump'' of sources
$<$ 4\arcsec.  

\begin{deluxetable}{ccccc}
\tablecolumns{5}
\tablewidth{0pc}
\tablecaption{\label{tab:wd1620com} Candidates around WD1620-391}
\tablehead{
\colhead{Candidate} & \colhead{R} & \colhead{PA} & \colhead{F110W} & \colhead{F160W}
}
\startdata
C1 & 2.22\arcsec$\pm$0.09 & 328.5$^\circ\pm$0.7 & 22.9 & 21.6 \\
C2 & 2.56\arcsec$\pm$0.13 & 262$^\circ\pm$5 & 22.9 & 21.8 \\
C3 & 3.10\arcsec$\pm$0.10 & 265$^\circ\pm$3 & 22.4 & 21.0 \\
C4 & 3.13\arcsec$\pm$0.14 & 141$^\circ\pm$1 & 23.9 & 23.0 \\
C5 & 3.24\arcsec$\pm$0.12 & 129.6$^\circ\pm$0.8 & 22.7 & 21.5 \\
C6 & 3.63\arcsec$\pm$0.17 & 27$^\circ\pm$2 & 22.5 & 21.2 \\
C7 & 3.91\arcsec$\pm$0.11 & 279$^\circ\pm$2 & 22.9 & 21.8 \\
\enddata
\end{deluxetable}

\begin{figure}
\plotone{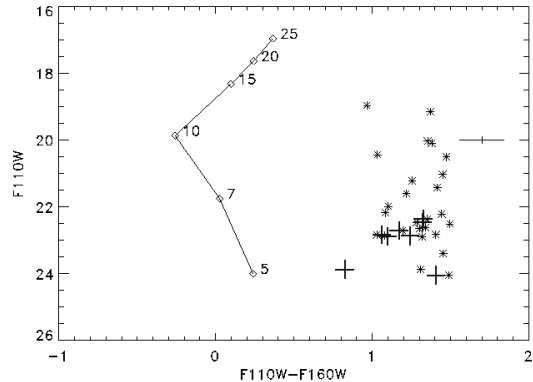}
\caption{\label{fig:cmd} Color magnitude diagram of sources near the white dwarf WD 1620-391.  Overplotted is an isochrone of 1 Gyr substellar models from
\citet{bsl03} convolved with HST filters at 12 pc. Thick crosses are sources $<$4\arcsec\ away.}
\end{figure}

Regardless, we have plotted all the detected sources in a CMD and compared them
to a predicted isochrone of substellar objects in Figure \ref{fig:cmd}.
The WDs age is $\sim$1 Gyr so we used the 1 Gyr 
models of \citet{bsl03} convolved with the HST filters.  There
are some candidates that are within 4\arcsec\ and who have colors consistent
within the errors to be a planetary candidate.  Table \ref{tab:wd1620com} lists the candidates, their magnitudes in F110W and
F160W.  Every one of the candidates would be $\sim$5-6 M$_{Jup}$ in mass if
associated.  This WDs proper motion is
$\sim$75 mas/yr in RA and $\sim$0 mas/yr in Dec \citep{hip}, 
so a second epoch image will be necessary in ruling out
any of these sources.

While we have two sets of observations for WD~1620-391 separated by six months,
our first image is not sensitive enough to conclusively detect
 any of our candidate companions.  Six stars were bright enough to use as a
background grid of reference stars compared to WD 1620-391's position.  
Of these six, five were distinct
point sources. The sixth appears to be extended,
either because it has a disk or because 
it is a binary.  When comparing the relative 
position between these presumably stationary objects in six months and WD 1620-391, we measured a change in RA of 204$\pm$10 mas and in Dec of 16$\pm$10 mas.
We derived the error based on the standard deviation of the individual measurements from the mean.  Taking into account WD 1620-391's parallax motion during
this period, one would expect a motion of 230 mas in RA and 28 mas in Dec
assuming WD 1620-391's reported parallax of 78.85 mas \citep{hip}.  Subtracting
this motion leaves 26$\pm$10 mas and 12$\pm$10 mas from the measured
motion with our reference stars, suggesting that we can 
detect common proper motion and common parallactic motion in a future epoch
with HST and these reference stars.  Since we have successfully proposed for HST time in Cycle 14
to follow up these candidates, we expect to have a long enough baseline to 
definitively determine if any of the candidates are physically associated.

\section{Limits to Companions}
\label{s4}
The main goal of this search was to detect candidate companions, but upper
 limits to the detection of such companions is also important for 
understanding the true nature of DAZ WDs, as well as the process of planet and
brown dwarf formation around intermediate mass stars.  To this end, in this 
Section we quantify our sensitivity to companions that could have
been detected, in order
to determine the frequency of high mass planets and brown dwarfs.

\subsection{Near-IR Photometry}
While direct imaging is most sensitive to companions $>$0.9\arcsec\,
unresolved companions could still be present for some of these targets.  In
order to rule out companions at separations where imaging or PSF subtraction
could not resolve them, we turn to the near-infrared fluxes of these objects
provided by near-IR photometry, such as from 2MASS \citep{cutri03}.  Looking
 in the near-IR can 
facilitate the discovery of cool objects around WDs \citep{probst82,zuckerman92,green00}. 

 Our strategy was to take model values reported in the literature,
generate predicted 2MASS J, H, and Ks  magnitudes by using the models
of \citet{bergeron95} and comparing  
J$_{th}$, H$_{th}$, and K$_{th}$ with the observed magnitudes of the WDs.  For our sample of
white dwarfs we took model values of T$_{eff}$, $\log{g}$, and the mass from
\citet{liebert04}, \citet{bergeron01}, and \citet{bragaglia95}. 

To compare the predicted magnitudes to those observed we took the difference 
of the predicted magnitdues in the 2MASS filter system J$_{th}$, H$_{th}$, and K$_{s(th)}$ and the observed magnitudes J, H, and K$_{s}$.
A significant positive value would indicate an excess due to either an unseen
companion or a dust disk, while a significant negative value would indicate an
anomalous paucity of flux.  While
we used the results of DSW05 for two of our white dwarfs for the rest of our targets we used the \citet{bergeron01} and \citet{bragaglia95} samples since they provide atmospheric parameters for the remaining five white
dwarfs.  In general, we compared J magnitudes since WD 2326+049 has an
infrared excess due to a dust disk at wavelengths longer than $\sim$1.6~\micron.  Excesses in J tend to be more sensitive because J band photometric errors
 are smaller in 2MASS.  For the rest of the targets we also checked to see if
there were excesses in any of the other bands or for other targets in the 
sample.  An excess was considered significant if it was greater than three
times the measured scatter of a sample and if it was present in more than one
filter.

We tested the accuracy of the three samples of WD parameters to reliably report a 3$\sigma$
excess limit.  We first examined the \citet{bergeron01} sample, which includes
WD 0208+395, WD 0245+541, WD 0243-025, and WD 1257+278.
Of the 150
white dwarfs we chose 146 of the sample that had reliable
photometry from \citet{bergeron01} 
 and converted their MKO magnitudes to 2MASS magnitudes
\footnote{http://www.ipac.caltech.edu/2mass/releases/allsky/doc/sec6\_4b.html}
to compare with our predicted magnitudes.  
 
We neglected any object
with an excess $>$ 3 $\sigma$ and recalculated the scatter in expected minus
observed magnitudes, repeating the process three times.  We ensured that the
median values of the differences were consistent with zero. 
 From the 146 WDs we find that the 1~$\sigma$ error in total
of J, H, and Ks are 0.04, 0.04, and 0.05 mag.  One important note is that \citet{bergeron01} used their JHK photometry to help fit several of the parameters 
that we used to generate our theoretical magnitudes, namely $\log{g}$ and 
T$_{eff}$.  For this reason we had to be more careful intrepeting these limits 
because it is possible the presence of a companion was ``fitted out''.
 In this 
case we are 
placing limits to what kind of excess would have been detected by the 
models, rather than extrapolating from the models and looking for excesses.
No objects in this sample showed a significant excess.

For WD 2326+049 and WD 1337+705 we took the sample
of \citet{liebert04}
which is a study of DA WDs from the Palomar-Green survey
of UV excess sources.  Of the 374
white dwarfs we chose the brightest 72 of the sample that had
 a J $<$ 15, had unambiguous sources in 2MASS, and had reliable
photometry, i.e those objects that had quality flags of A or B in the 2MASS
point source catalogue for their J magnitudes.  After determining the standard deviation of the sample, we found that 1$\sigma$\
errors for the sample in the J, H, and K$_s$ filters were 0.07, 0.10, and 0.15 mag, respectively.  Further details of the \citet{liebert04} sample are presented in DSW05.

 For WD 1620-391, we needed to use
the sample in \citet{bragaglia95}, using $\sim$35 of the 50 WDs modeled in that
work.  We again picked WDs with V$<$15, reliable 2MASS positions, and 
reliable photometry in the three bands.
  Six  white dwarfs had poor photometry or incorrect distance moduli,
 but these errors were corrected. 
  The final errors were 
calculated, resulting in 1$\sigma$ errors of 0.09, 0.08, 0.15 mag 
for J, H, and Ks
respectively.  Two WDs remained with significant excess, WD 1042-690, and 
WD 1845+019.  WD 1042-690 is a known binary system with a dM companion, and 
WD 1845+019 does not currently seem to be a candidate for an excess.  
However, its 
position in both the POSS and 2MASS plates based on the position given
by \citet{lanning00} shows that it is
blended with another point source.  
Inspection of the POSS and 2MASS plates leaves it ambiguous whether this 
barely resolved object (separcdation $\sim$3\arcsec) is co-moving or not, so we
mark this as a potential common proper motion WD/dM pair. 

Table \ref{tab:phot} shows the expected 2MASS magnitudes
based on the model values,
 and the observed magnitudes of our target white dwarfs.  All of our targets
fall within 1-2$\sigma$ of our expected values for all three filters, with the
exception of WD 2326+049, as mentioned above.

Since none of our targets have significant excesses, we can use the
 3$\sigma$ limits in J to place upper limits to 
unresolved sources.  We took the predicted J magnitudes
from substellar atmosphere models, corrected for distance modulus,
calculated the excess, and compared it to
our sensitivity limit \citep{baraffe98,baraffe03}.  
Table \ref{tab:sens} shows the unresolved companion
upper limits for each target.  Any companion with a mass beyond the hydrogen
burning limit would have been detected for all of the target WDs.
\begin{deluxetable}{ccccccc}
\tablecolumns{7}
\tablewidth{0pc}
\tablecaption{\label{tab:phot} Comparison of Predicted vs. 2MASS Photometry}
\tablehead{
\colhead{WD} & \colhead{J$_{th}$} & \colhead{H$_{th}$} & \colhead{K$_{s(th)}$}
 & \colhead{J} & \colhead{H} & \colhead{K$_{s}$} }
 \startdata
0208+396 & 13.74 & 13.61 & 13.57 & 13.76 & 13.66 & 13.61 \\
0243-026 & 14.65 & 14.49 & 14.43 & 14.67 & 14.50 & 14.49 \\
0245+541 & 13.86 & 13.61 & 13.47 & 13.86 & 13.67 & 13.58 \\
1257+278 & 14.95 & 14.89 & 14.88 & 14.95 & 14.92 & 14.89  \\
1337+701 & 13.23 & 13.36 & 13.41 & 13.25 & 13.36 & 13.45 \\
1620-391 & 11.53 & 11.66 & 11.74 & 11.58 & 11.71 & 11.77 \\
2326+049 & 13.13 & 13.19 & 13.22 & 13.13 & 13.08 & 12.69 \\
\enddata
\end{deluxetable}

\subsection{Imaging}
\label{s4.2}

\citet{schneider03} showed
 a reliable way to determine sensitivity of an observation with
NICMOS, given the stability of the instrument.  Artificial ``companions'' are
generated with the HST PSF simulation software TINYTIM \footnote{http://www.stsci.edu/software/tinytim/tinytim.html} and scaled until they are recovered.
  These companions are 
 inserted into the observations and used to gauge sensitivity.  We adopted
this strategy for our data as well.  An implant was placed in the images.
  Two difference
images were created following our procedure of PSF subtraction
and then rotated and combined for maximum signal to noise.
Sample images were looked
at by eye as a second check that the dimmest implants could be recovered.
The implants were normalized so that their total flux was equal to
1 DN/s.   The normalized value was converted to a flux in Jy or a Vega magnitude by 
multiplying by the correct photometry constants given by the NICMOS Data Handbook.   
We considered an implant recovered if its scaled
flux in a given aperture had a S/N of 5.   

\begin{figure}
\plotone{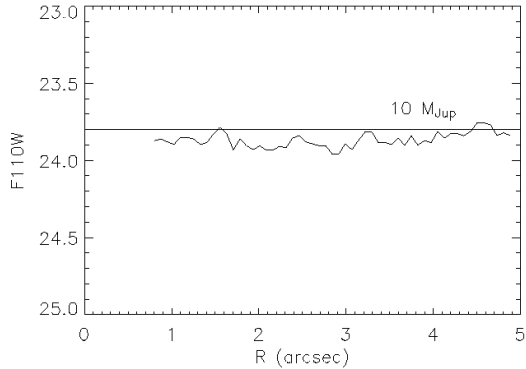}
\caption{\label{fig:wdsens} Sensitivity at 5$\sigma$ to point sources in F110W
around the WD WD 0208+395.  The WDs F110W magnitude is $\sim$13.8, giving a contrast
of 10 magnitudes at 1\arcsec.  Overplotted is the magnitude of a 10 M$_{Jup}$
planet 3.2 Gyr old at the distance of WD 0208+395 from the models of \citet{bsl03}.}
\end{figure}

 For our Gemini data, we
used the PSF of WD 2326+049 as a reference for the implant.  The implant was normalized to a peak pixel value of one.  Scaled versions of the implants were then 
used to determine the final image's 
sensitivity to objects at a S/N of 10, since siginificant
 flux from the PSF remained at separations $<$ 1\arcsec.  The relative flux of the implant with respect
to the host star was measured and a corresponding MKO
 H magnitude was derived from the 2MASS H magnitude to give a final apparent
magnitude sensitivity.  For our Gemini images we checked sensitivity starting 
at a distance of $\sim$3 times the FWHM of WD 2326+049, or  0.22\arcsec,
 out to 1\arcsec.

Figure \ref{fig:wdsens} shows an example of the azimuthally averaged 5 $\sigma$
sensitivity for WD 0208+395.  In order to determine the total age of WD 0208+396 as well as each other system, we took each inferred mass and 
derived an initial mass by the relation $10.4\ln[(M_{WD}/\Msun)/0.49] \Msun$, the results of which are given
in Table \ref{tab:targs} \citep{wood92}.  The mass then gave a main sequence lifetime given by 10 t$_{MS}^{-2.5}$~Gyr, which gave a total age when coupled with the inferred 
cooling time from the same models used for our 2MASS photometry.
  With a total estimated system age of $\sim$3
Gyr for WD 0208+395, we overplotted the lowest companion mass detectable, using
the models of \citet{bsl03}.  These models differ slightly from the 
models of \citet{baraffe03}, used for our 2MASS excess limits.  The \citet{bsl03} models tend to predict dimmer near-IR magnitudes for
the planetary mass objects, but converge with the \citet{baraffe03} models
for higher masses.  It is therefore possible that we are sensitive to objects
$\sim$1-2 M$_{Jup}$ less massive if the \citet{baraffe03} models are 
correct.  Table \ref{tab:sens}
has the mass limits for each WD for separations $>$ 0.9\arcsec.  

\section{Discussion}
\label{s5}

We can use Table \ref{tab:sens} and the results of our excess limits to draw 
some broad conclusions from this search.  The combination of the 2MASS excess
determinations and the HST imaging create the most sensitive search for 
planets around WDs to date.  The sensitivity achieved could easily have detected
an object $>$ 10 M$_{Jup}$ at separations $>$ 30.6 AU, with the closest
detection possible at 9.3 AU.  Taking into account that any primordial 
companions' semi-major axis would have expanded by a factor of $M_i/M_f$, we
can infer the closest primordial separation these objects would have had if 
they had been
detected.  Taking the values of Table \ref{tab:targs} for M$_i$ and M$_f$ and 
using our minimum projected angular separation, we 
find that any object that formed at $>$10 AU could have been
detected, assuming that there were no forces that retarded expansion.  Forces
that could retard expansion would be
due to tidal interactions with the giant star.  However, this effect should be minimal at initial distances of 10 AU \citep{rasio96}. 

We can also make some initial comments about the origin of DAZ white dwarfs.
Given the upper limits on unresolved companions, we can infer the plausibility
of one of the possible explanations for the DAZ phenomenon.  The problems with
ISM accretion have been documented extensively in the work of \citet{zuckerman03} and \citet{aannestad93}.  In \citet{zuckerman03} they noted that a large fraction of 
DA/dM objects had metal absorption lines in their atmospheres, and inferred
that other DAZs may be the result of unseen companions.  If this scenario is 
true, then for each of these objects, the maximum companion mass plausible
is $<$70 M$_{Jup}$.  

If the explanation
for DAZs is due to close brown dwarf companions, the frequency of DAZs is at 
odds with the frequency of DAZs one would predict based on radial velocity
surveys.  These surveys find that $\sim$0.5\% of stars have brown dwarfs with
semi-major axes $<$3~AU \citep{marcy00}.  One would expect 0.5\% or less of field DAs to be
DAZs based on the radial velocity result.  The only possible counter explanation is that brown dwarf
 formation at these radii is $\sim$40 times more efficient for
higher mass main sequence stars.
Radial velocity surveys of G giants are too young
to reliably estimate the fraction of brown dwarf companions in orbits wider than $\sim$1 AU, but none have yet been found in $\sim$100 stars
\citep{sato03}.  

\begin{deluxetable*}{lcccc}
\tablecolumns{5}
\tablewidth{0pc}
\tablecaption{\label{tab:sens} Upper limits to Companions}
\tablehead{
\colhead{WD} & \colhead{Excess Limit (m$_{J}$)} & \colhead{Mass} & \colhead{Sensitivity $>$ 0.9\arcsec} & \colhead{Mass} \\
 & \colhead{(J)} & \colhead{(M$_{Jup}$)} & \colhead{(F110W)} & \colhead{(M$_{Jup}$)}}
\startdata
0208+396 & 16.2 & 48 & 23.9 & 10 \\
0243-026 & 17.5 & 51 & 24.1 & 10 \\
0245+541 & 16.2 & 53 & 23.5 & 18 \\
1257+278 & 17.0 & 40 & 23.8 & 14 \\
1337+701 & 14.9 & 70 & 23.4 & 14 \\
1620-391 & 12.9 & 61 & 22.9 & 7  \\
2326+049 & 14.8 & 39 & 23.3 & 6  \\
\enddata
\end{deluxetable*}

We can compare our results with those of radial velocity surveys.  By comparing
both results we can look at predictions for the frequency of massive
planets around
a random sample of stars and around stars that possess planetary systems.
Since the numbers are small, we will merely look at percentages and assume that
they are constant as a function of distance and central stellar mass, 
clearly naive assumptions.  Since 5\% of field stars have
planetary systems, we need to estimate how many would have planets
massive enough to be detectable by our observations.  Of the
118 known planetary systems in orbit around solar type stars, $\sim$6
 have companions with M$\sin i$ $>$ 10 M$_{Jup}$\footnote{http://www.obspm.fr/encycl/encycl.html}.  The frequency of such planets amongst stars already
bearing one or more planets is then $\sim$5\%, leading to an overall 
probability of 0.25\% of all field stars possessing a planet that we could 
have detected.  Assuming Poisson statistics,
to have a 50\% chance
 at detecting one or two
planets would require a sample of 400 WDs with ages $\sim$3 Gyr.
The limit sensitive radial velocity studies have on
A stars can be partially circumvented
by searching G giants for radial velocity variations \citep{sato03}.  G giants
are typically intermediate mass stars, although field giants tend to have
larger uncertainties in their mass compared to the main sequence stars in 
other radial velocity surveys.  As of the results published in \citet{sato03},
one planetary object with M$\sin{i}$=6-10 M$_{Jup}$ and semi-major axis
$\sim$1~AU had been detected in a 
sample of $\sim$100 targets.
The implied frequency of $\sim$1\% would mean a slightly more favorable chance
to find one planet in a sample of $\sim$100 WDs.
If DAZs do not preferentially harbor planetary systems, it will be a long search if we only focus on them.  Any search should include DAZs, but also focus 
on a larger sample.  

Let us now consider the possibility that DAZs do preferentially harbor planetary systems, and based on our detection limits estimate 
how many DAZs would need to be 
observed.  Since we could detect $>$ 10~M$_{Jup}$ objects and $\sim$5\% of
field stars with planetary systems have objects that massive, we can infer that
5\% of DAZs could have planets that could have been detected.
If DAZs (and also
DZs or helium white dwarfs with metal absorption) are indeed 
good markers for planetary systems, one would need a sample of
20 WDs to have a 50\% chance to 
detect a massive planet.  To date $\sim$34 DAZs are known.
Currently the estimated fraction of 
apparent single WDs that are DAZs is $\sim$20\%. If they all harbor planets, this estimated fraction implies a much higher
frequency of planets than that measured by radial velocity surveys.  However,
radial velocity 
surveys are starting to detect longer period systems, which may have a 
higher frequency of formation and better represent the type of population
that would cause a DAZ \citep{jones02}.

There are then two approaches to continuing the search--increasing the sample
size and increasing the sensitivity of a search.  In the short term a large sample of WDs must be observed, since the 
probable frequency of massive planets among WDs that harbor a planetary system
is small.  
  Future observatories such as the James Webb
Space Telescope should have an easier time detecting Jovian and sub-Jovian 
planets, which will hopefully resolve the origin of DAZs.  Such future observations will determine whether DAZs are 
ultimately useful for planetary studies, including spectroscopy.

The discovery of candidate planetary mass companions demonstrates that this
limited survey was sensitive to planets.  These results
 show that if massive planets were present
around these WDs we would have detected them.  Even with a small sample, limits
can be placed on the frequency of massive planets in orbit around stars more
massive than the Sun, and begin to observationally address the question of 
planet formation efficiency vs. spectral type.  Ideally, the next step would be
to be to expand the sample of WDs studied and to probe to lower masses, where the planetary mass function peaks ($\sim$1 M$_{Jup}$).  High spatial resolution and sensitivity missions like JWST would most
likely be able to detect such objects around nearby WDs.   

\acknowledgements
We would like to gratefully acknowledge Al Shultz and Glenn Schneider for
helpful conversations about coronagraphy with NICMOS, and Chad Trujillo and Joe
Jensen for critical help with the inner workings of Altair and the reduction
of Altair imaging data.

Based on observations made with the NASA/ESA Hubble Space Telescope, obtained
 at the Space Telescope Science Institute, which is operated by the 
Association of Universities for Research in Astronomy, Inc., under NASA
contract NASÊ5-26555. These observations are associated with program \#9834.
Also based on observations obtained at the Gemini Observatory,
which is operated by the
Association of Universities for Research in Astronomy, Inc.,
under a cooperative agreement
with the NSF on behalf of the Gemini partnership: the National
 Science Foundation (United
States), the Particle Physics and Astronomy Research Council
(United Kingdom), the
National Research Council (Canada), CONICYT (Chile), the Australian
 Research Council
(Australia), CNPq (Brazil) and CONICET (Argentina). Near-IR Photometry obtained as part of the Two Micron All Sky Survey (2MASS), a joint project of the University
of Massachusetts and the Infrared Processing and Analysis Center/California
Institute of Technology, funded by the National Aeronautics and Space
Administration and the National Science Foundation.  S.S. also acknowledges
funding under the Pennsylvania State University Astrobiology Research Consortium (PSARC).

\bibliography{g29bib}

\begin{thebibliography}{61}
\expandafter\ifx\csname natexlab\endcsname\relax\def\natexlab#1{#1}\fi

\bibitem[{{Aannestad} {et~al.}(1993){Aannestad}, {Kenyon}, {Hammond}, \&
  {Sion}}]{aannestad93}
{Aannestad}, P.~A., {Kenyon}, S.~J., {Hammond}, G.~L., \& {Sion}, E.~M. 1993,
  \aj, 105, 1033

\bibitem[{{Alcock} {et~al.}(1986){Alcock}, {Fristrom}, \&
  {Siegelman}}]{alcock86}
{Alcock}, C., {Fristrom}, C.~C., \& {Siegelman}, R. 1986, \apj, 302, 462

\bibitem[{{Althaus} \& {Benvenuto}(1998)}]{althaus98}
{Althaus}, L.~G. \& {Benvenuto}, O.~G. 1998, \mnras, 296, 206

\bibitem[{{Bakos} {et~al.}(2002){Bakos}, {Sahu}, \& {N{\' e}meth}}]{bakos02}
{Bakos}, G.~{\' A}., {Sahu}, K.~C., \& {N{\' e}meth}, P. 2002, \apjs, 141, 187

\bibitem[{{Baraffe} {et~al.}(1998){Baraffe}, {Chabrier}, {Allard}, \&
  {Hauschildt}}]{baraffe98}
{Baraffe}, I., {Chabrier}, G., {Allard}, F., \& {Hauschildt}, P.~H. 1998, \aap,
  337, 403

\bibitem[{{Baraffe} {et~al.}(2003){Baraffe}, {Chabrier}, {Barman}, {Allard}, \&
  {Hauschildt}}]{baraffe03}
{Baraffe}, I., {Chabrier}, G., {Barman}, T.~S., {Allard}, F., \& {Hauschildt},
  P.~H. 2003, \aap, 402, 701

\bibitem[{{Bergeron} {et~al.}(2001){Bergeron}, {Leggett}, \&
  {Ruiz}}]{bergeron01}
{Bergeron}, P., {Leggett}, S.~K., \& {Ruiz}, M.~T. 2001, \apjs, 133, 413

\bibitem[{{Bergeron} {et~al.}(1995){Bergeron}, {Wesemael}, {Lamontagne},
  {Fontaine}, {Saffer}, \& {Allard}}]{bergeron95}
{Bergeron}, P., {Wesemael}, F., {Lamontagne}, R., {Fontaine}, G., {Saffer},
  R.~A., \& {Allard}, N.~F. 1995, \apj, 449, 258

\bibitem[{{Bragaglia} {et~al.}(1995){Bragaglia}, {Renzini}, \&
  {Bergeron}}]{bragaglia95}
{Bragaglia}, A., {Renzini}, A., \& {Bergeron}, P. 1995, \apj, 443, 735

\bibitem[{{Burgasser} {et~al.}(2003){Burgasser}, {Kirkpatrick}, {McElwain},
  {Cutri}, {Burgasser}, \& {Skrutskie}}]{burgasser03}
{Burgasser}, A.~J., {Kirkpatrick}, J.~D., {McElwain}, M.~W., {Cutri}, R.~M.,
  {Burgasser}, A.~J., \& {Skrutskie}, M.~F. 2003, \aj, 125, 850

\bibitem[{{Burleigh} {et~al.}(2002){Burleigh}, {Clarke}, \&
  {Hodgkin}}]{burleigh02}
{Burleigh}, M.~R., {Clarke}, F.~J., \& {Hodgkin}, S.~T. 2002, \mnras, 331, L41

\bibitem[{{Burrows} {et~al.}(2004){Burrows}, {Sudarsky}, \&
  {Hubeny}}]{burrows04}
{Burrows}, A., {Sudarsky}, D., \& {Hubeny}, I. 2004, \apj, 609, 407

\bibitem[{{Burrows} {et~al.}(2003){Burrows}, {Sudarsky}, \& {Lunine}}]{bsl03}
{Burrows}, A., {Sudarsky}, D., \& {Lunine}, J.~I. 2003, \apj, 596, 587

\bibitem[{{Cutri} {et~al.}(2003){Cutri}, {Skrutskie}, {van Dyk}, {Beichman},
  {Carpenter}, {Chester}, {Cambresy}, {Evans}, {Fowler}, {Gizis}, {Howard},
  {Huchra}, {Jarrett}, {Kopan}, {Kirkpatrick}, {Light}, {Marsh}, {McCallon},
  {Schneider}, {Stiening}, {Sykes}, {Weinberg}, {Wheaton}, {Wheelock}, \&
  {Zacarias}}]{cutri03}
{Cutri}, R.~M., {et al.} 2003, VizieR Online Data Catalog, 2246, 0

\bibitem[{{Debes} {et~al.}(2005{\natexlab{a}}){Debes}, {Ge}, \&
  {Ftaclas}}]{debes05c}
{Debes}, J.~H., {Ge}, J., \& {Ftaclas}, C. 2005{\natexlab{a}}, in preparation

\bibitem[{{Debes} \& {Sigurdsson}(2002)}]{debes02}
{Debes}, J.~H. \& {Sigurdsson}, S. 2002, \apj, 572, 556

\bibitem[{{Debes} {et~al.}(2005{\natexlab{b}}){Debes}, {Sigurdsson}, \&
  {Woodgate}}]{debes05a}
{Debes}, J.~H., {Sigurdsson}, S., \& {Woodgate}, B. 2005{\natexlab{b}}, (ApJ,
  submitted)

\bibitem[{{Delfosse} {et~al.}(1998){Delfosse}, {Forveille}, {Mayor}, {Perrier},
  {Naef}, \& {Queloz}}]{delfosse98}
{Delfosse}, X., {Forveille}, T., {Mayor}, M., {Perrier}, C., {Naef}, D., \&
  {Queloz}, D. 1998, \aap, 338, L67

\bibitem[{{Dupuis} {et~al.}(1992){Dupuis}, {Fontaine}, {Pelletier}, \&
  {Wesemael}}]{dupuis92}
{Dupuis}, J., {Fontaine}, G., {Pelletier}, C., \& {Wesemael}, F. 1992, \apjs,
  82, 505

\bibitem[{{Dupuis} {et~al.}(1993{\natexlab{a}}){Dupuis}, {Fontaine},
  {Pelletier}, \& {Wesemael}}]{dupuis93a}
---. 1993{\natexlab{a}}, \apjs, 84, 73

\bibitem[{{Dupuis} {et~al.}(1993{\natexlab{b}}){Dupuis}, {Fontaine}, \&
  {Wesemael}}]{dupuis93b}
{Dupuis}, J., {Fontaine}, G., \& {Wesemael}, F. 1993{\natexlab{b}}, \apjs, 87,
  345

\bibitem[{{Ford} {et~al.}(2001){Ford}, {Havlickova}, \& {Rasio}}]{ford01}
{Ford}, E.~B., {Havlickova}, M., \& {Rasio}, F.~A. 2001, Icarus, 150, 303

\bibitem[{{Forveille} {et~al.}(2004){Forveille}, {S{\' e}gransan}, {Delorme},
  {Mart{\'{\i}}n}, {Delfosse}, {Acosta-Pulido}, {Beuzit}, {Manchado}, {Mayor},
  {Perrier}, \& {Udry}}]{forveille04}
{Forveille}, T., {et al.} 2004, \aap, 427, L1

\bibitem[{{Fraquelli} {et~al.}(2004){Fraquelli}, {Schultz}, {Bushouse}, {Hart},
  \& {Vener}}]{fraquelli04}
{Fraquelli}, D.~A., {Schultz}, A.~B., {Bushouse}, H., {Hart}, H.~M., \&
  {Vener}, P. 2004, \pasp, 116, 55

\bibitem[{{Gianninas} {et~al.}(2004){Gianninas}, {Dufour}, \&
  {Bergeron}}]{gianninas04}
{Gianninas}, A., {Dufour}, P., \& {Bergeron}, P. 2004, \apjl, 617, L57

\bibitem[{{Gladman}(1993)}]{gladman95}
{Gladman}, B. 1993, Icarus, 106, 247

\bibitem[{{Grady} {et~al.}(1999){Grady}, {Woodgate}, {Bruhweiler}, {Boggess},
  {Plait}, {Lindler}, {Clampin}, \& {Kalas}}]{grady99}
{Grady}, C.~A., {Woodgate}, B., {Bruhweiler}, F.~C., {Boggess}, A., {Plait},
  P., {Lindler}, D.~J., {Clampin}, M., \& {Kalas}, P. 1999, \apjl, 523, L151

\bibitem[{{Graham} {et~al.}(1990){Graham}, {Matthews}, {Neugebauer}, \&
  {Soifer}}]{graham90}
{Graham}, J.~R., {Matthews}, K., {Neugebauer}, G., \& {Soifer}, B.~T. 1990,
  \apj, 357, 216

\bibitem[{{Green} {et~al.}(2000){Green}, {Ali}, \& {Napiwotzki}}]{green00}
{Green}, P.~J., {Ali}, B., \& {Napiwotzki}, R. 2000, \apj, 540, 992

\bibitem[{{Griffin} {et~al.}(2000){Griffin}, {David}, \&
  {Verschueren}}]{griffin00}
{Griffin}, R.~E.~M., {David}, M., \& {Verschueren}, W. 2000, \aaps, 147, 299

\bibitem[{{Hawley} {et~al.}(2002){Hawley}, {Covey}, {Knapp}, {Golimowski},
  {Fan}, {Anderson}, {Gunn}, {Harris}, {Ivezi{\' c}}, {Long}, {Lupton},
  {McGehee}, {Narayanan}, {Peng}, {Schlegel}, {Schneider}, {Spahn}, {Strauss},
  {Szkody}, {Tsvetanov}, {Walkowicz}, {Brinkmann}, {Harvanek}, {Hennessy},
  {Kleinman}, {Krzesinski}, {Long}, {Neilsen}, {Newman}, {Nitta}, {Snedden}, \&
  {York}}]{hawley02}
{Hawley}, S.~L., {et al.}  2002, \aj, 123, 3409

\bibitem[{{Hill}(1886)}]{hill86}
{Hill}, G.~W. 1886, Acta Mathematica, 8, 1

\bibitem[{{Holberg} {et~al.}(1995){Holberg}, {Bruhweiler}, \&
  {Andersen}}]{holberg95}
{Holberg}, J.~B., {Bruhweiler}, F.~C., \& {Andersen}, J. 1995, \apj, 443, 753

\bibitem[{{Holland} {et~al.}(1998){Holland}, {Greaves}, {Zuckerman}, {Webb},
  {McCarthy}, {Coulson}, {Walther}, {Dent}, {Gear}, \& {Robson}}]{holland98}
{Holland}, W.~S., {et al.}  1998, \nat, 392, 788

\bibitem[{{Hutchings} {et~al.}(2004){Hutchings}, {Stoesz}, {Veran}, \&
  {Rigaut}}]{hutchings04}
{Hutchings}, J.~B., {Stoesz}, J., {Veran}, J.-P., \& {Rigaut}, F. 2004, \pasp,
  116, 154

\bibitem[{{Ignace}(2001)}]{ignace01}
{Ignace}, R. 2001, \pasp, 113, 1227

\bibitem[{{Jayawardhana} {et~al.}(1998){Jayawardhana}, {Fisher}, {Hartmann},
  {Telesco}, {Pina}, \& {Fazio}}]{rayjay99}
{Jayawardhana}, R., {Fisher}, S., {Hartmann}, L., {Telesco}, C., {Pina}, R., \&
  {Fazio}, G. 1998, \apjl, 503, L79+

\bibitem[{{Jeans}(1924)}]{jeans24}
{Jeans}, J.~H. 1924, \mnras, 85, 2

\bibitem[{{Jones} {et~al.}(2002){Jones}, {Paul Butler}, {Marcy}, {Tinney},
  {Penny}, {McCarthy}, \& {Carter}}]{jones02}
{Jones}, H.~R.~A., {Paul Butler}, R., {Marcy}, G.~W., {Tinney}, C.~G., {Penny},
  A.~J., {McCarthy}, C., \& {Carter}, B.~D. 2002, \mnras, 337, 1170

\bibitem[{{Jura}(2003)}]{jura03}
{Jura}, M. 2003, \apjl, 584, L91

\bibitem[{{Lada} \& {Lada}(2003)}]{lada03}
{Lada}, C.~J. \& {Lada}, E.~A. 2003, \araa, 41, 57

\bibitem[{{Lanning}(2000)}]{lanning00}
{Lanning}, H.~H. 2000, VizieR Online Data Catalog, 2231, 0

\bibitem[{{Liebert} {et~al.}(2005){Liebert}, {Bergeron}, \&
  {Holberg}}]{liebert04}
{Liebert}, J., {Bergeron}, P., \& {Holberg}, J.~B. 2005, \apjs, 156, 47

\bibitem[{{Marcy} \& {Butler}(2000)}]{marcy00}
{Marcy}, G.~W. \& {Butler}, R.~P. 2000, \pasp, 112, 137

\bibitem[{{Parriott} \& {Alcock}(1998)}]{parriott98}
{Parriott}, J. \& {Alcock}, C. 1998, \apj, 501, 357

\bibitem[{{Patterson} {et~al.}(1991){Patterson}, {Zuckerman}, {Becklin},
  {Tholen}, \& {Hawarden}}]{patterson91}
{Patterson}, J., {Zuckerman}, B., {Becklin}, E.~E., {Tholen}, D.~J., \&
  {Hawarden}, T. 1991, \apj, 374, 330

\bibitem[{{Perryman} {et~al.}(1997){Perryman}, {Lindegren}, {Kovalevsky},
  {Hoeg}, {Bastian}, {Bernacca}, {Cr{\' e}z{\' e}}, {Donati}, {Grenon}, {van
  Leeuwen}, {van der Marel}, {Mignard}, {Murray}, {Le Poole}, {Schrijver},
  {Turon}, {Arenou}, {Froeschl{\' e}}, \& {Petersen}}]{hip}
{Perryman}, M.~A.~C., {et al.}  1997, \aap, 323, L49

\bibitem[{{Probst} \& {Oconnell}(1982)}]{probst82}
{Probst}, R.~G. \& {Oconnell}, R.~W. 1982, \apjl, 252, L69

\bibitem[{{Rasio} {et~al.}(1996){Rasio}, {Tout}, {Lubow}, \& {Livio}}]{rasio96}
{Rasio}, F.~A., {Tout}, C.~A., {Lubow}, S.~H., \& {Livio}, M. 1996, \apj, 470,
  1187

\bibitem[{{Sato} {et~al.}(2003){Sato}, {Ando}, {Kambe}, {Takeda}, {Izumiura},
  {Masuda}, {Watanabe}, {Noguchi}, {Wada}, {Okada}, {Koyano}, {Maehara},
  {Norimoto}, {Okada}, {Shimizu}, {Uraguchi}, {Yanagisawa}, \&
  {Yoshida}}]{sato03}
{Sato}, B., {et al.}  2003, \apjl, 597, L157

\bibitem[{{Schneider} \& {Silverstone}(2003)}]{schneider03}
{Schneider}, G. \& {Silverstone}, M.~D. 2003, in High-Contrast Imaging for
  Exo-Planet Detection. Edited by Alfred B. Schultz. Proceedings of the SPIE,
  Volume 4860, pp. 1-9 (2003)., 1--9

\bibitem[{{Setiawan} {et~al.}(2005){Setiawan}, {Rodmann}, {da Silva}, {Hatzes},
  {Pasquini}, {von der Luehe}, {de Medeiros}, {Doellinger}, \&
  {Girardi}}]{setiawan05}
{Setiawan}, J., {et al.} 2005, ArXiv Astrophysics e-prints

\bibitem[{{Sigurdsson} {et~al.}(2003){Sigurdsson}, {Richer}, {Hansen},
  {Stairs}, \& {Thorsett}}]{sigurdsson03}
{Sigurdsson}, S., {Richer}, H.~B., {Hansen}, B.~M., {Stairs}, I.~H., \&
  {Thorsett}, S.~E. 2003, Science, 301, 193

\bibitem[{{Tremaine}(1993)}]{tremaine92}
{Tremaine}, S. 1993, in ASP Conf. Ser. 36: Planets Around Pulsars, 335--344

\bibitem[{{van Altena} {et~al.}(2001){van Altena}, {Lee}, \&
  {Hoffleit}}]{vanaltena95}
{van Altena}, W.~F., {Lee}, J.~T., \& {Hoffleit}, E.~D. 2001, VizieR Online
  Data Catalog, 1238, 0

\bibitem[{{Weinberger} {et~al.}(1999){Weinberger}, {Becklin}, {Schneider},
  {Smith}, {Lowrance}, {Silverstone}, {Zuckerman}, \& {Terrile}}]{weinberger99}
{Weinberger}, A.~J., {Becklin}, E.~E., {Schneider}, G., {Smith}, B.~A.,
  {Lowrance}, P.~J., {Silverstone}, M.~D., {Zuckerman}, B., \& {Terrile}, R.~J.
  1999, \apjl, 525, L53

\bibitem[{{Wolszczan} \& {Frail}(1992)}]{wolszczan92}
{Wolszczan}, A. \& {Frail}, D.~A. 1992, \nat, 355, 145

\bibitem[{{Wood}(1992)}]{wood92}
{Wood}, M.~A. 1992, \apj, 386, 539

\bibitem[{{Zuckerman} \& {Becklin}(1992)}]{zuckerman92}
{Zuckerman}, B. \& {Becklin}, E.~E. 1992, \apj, 386, 260

\bibitem[{{Zuckerman} {et~al.}(2003){Zuckerman}, {Koester}, {Reid}, \& {H{\"
  u}nsch}}]{zuckerman03}
{Zuckerman}, B., {Koester}, D., {Reid}, I.~N., \& {H{\" u}nsch}, M. 2003, \apj,
  596, 477

\bibitem[{{Zuckerman} \& {Reid}(1998)}]{zuckerman98}
{Zuckerman}, B. \& {Reid}, I.~N. 1998, \apjl, 505, L143

\end{thebibliography}
\bibliographystyle{apj}

\end{document}